\documentclass{article}
\usepackage{epsfig}
\begin{document}
\title{A nanomechanical resonator shuttling single electrons at radio
frequencies}
\author{A. Erbe$^*$, C. Weiss$^{**}$, W. Zwerger$^{**}$,
and R.H. Blick$^*$\\ \\ Center for NanoScience and Sektion Physik,
Ludwig-Maximilians-Universit\"at, 
\\ $^{*}$Geschwister-Scholl-Platz 1, 80539 M\"unchen, Germany 
\\ $^{**}$Theresienstr.~37,
80333 M\"unchen, Germany \\}

\date{November 8, 2000}

\maketitle

\begin{abstract}
We observe transport of electrons through a metallic
island on the tip of a nanomechanical pendulum.
The resulting tunneling current shows distinct features
corresponding to the discrete mechanical eigenfrequencies of the pendulum.
We report on 
measurements covering the temperature range from 300~K down to 4.2~K.
We explain the $I$-$V$ curve, which 
unexpectedly differs from previous theoretical predictions, 
with model
calculations based on a Master equation approach.
\end{abstract}

One of the traditional experiments in the electrodynamics class
is set up by two large capacitor plates and a metallized ball suspended in
between the plates.  
Applying a constant voltage of several 100~V across
the plates leads
to the onset of periodic charge transfer by the ball bouncing back and forth,
similar to a classical bell~\cite{benjamin1895}.
The number of electrons transferred by the metallized ball in each revolution
naturally depends on the volume of the metal, but can be estimated to be of
the order of $10^{10}$. At an oscillation frequency of some 10~Hz up into 
the audible kHz-range this gives a typical current of $1 - 10~\mu$A.

The question arising is whether such an experiment can be performed on the
microscopic level in order to obtain a transfer not of a multitude 
but of only one
electron per cycle of operation at frequencies of some 100~MHz. 
Indeed this can be achieved by
simply scaling down the setup and applying a
nanomechanical  resonator. In recent experiments \cite{mce00} the importance of the
excitation of mechanical
modes for electronic transport through single fullerenes was discussed.

Here we want
to present our results on shrinking the mechanical electron shuttle to
submicron dimensions by 
integration of an electron island into a nanomechanical resonator 
functioning as an electromechanical transistor (EMT). 
The clear advantages are the increased speed of operation and
the reduction of the transfer rate, allowing to count electrons one by one. 
A similar combination was proposed
theoretically by Gorelik~{\it et al.}~\cite{gorelik98:4526} for metallic
particles, which are connected to the reservoirs by elastically
deformable organic molecular links. The main difference to common
single electron transistor 
devices is the
fact that only one tunneling barrier is open at a certain time. This
leads to an exponential suppression of cotunneling effects and thus
increases the accuracy of current transport. Detailed calculations for theoretical limits 
of the accuracy, which were presented elsewhere~\cite{weiss99:97}, 
show that these devices will allow 
measuring quantum fluctuations. In the present work we report on
measurements of transport of electrons through a
nanomechanical electron shuttle. The $I$-$V$ characteristics
which unexpectedly
differs from theoretical predictions~\cite{weiss99:97} can be
explained by taking into account the driving voltage.

An important feature of the device is the possibility to effectively
modulate the tunneling
rate onto and off the electron island given by the large speed of
operation~($f \sim 100$~MHz). This basically enables
to mechanically filter and select electrons passing through the
electromechanical circuit by simply adjusting the tunneling rate~$\Gamma$.
Additionally, the device is a tool to regularize the stochastic tunneling process
through a well defined geometry. 
Another advantage of nanomechanical resonators, machined out of
Silicon-On-Insulator~(SOI) 
material, is their insensitivity to thermal and electrical shocks
as has been shown in their application for
electrometry~\cite{cleland96:2653,kroemmer00:}. 
This and the high speed of operation enable direct integration in
filter applications~\cite{sensor}. We have already 
demonstrated, that a nanomechanical tunneling contact, which operates at radio
frequencies, can be built out of SOI substrates~\cite{erbe98}. 
The tunneling process in turn is very sensitive to changes of the
environmental conditions, thus its use in sensor applications. 


The nano-electromechanical electron shuttle is machined out of SOI-material. 
In order to avoid mechanical defects in the structure so-called
``Smart Cut''~\cite{smartcut} is used rather than SIMOX~\cite{simox}.
The fabrication process is divided into two steps: First the metallic
leads (made out of Au) and the etch mask (made out of Al) are defined
by optical and electron beam lithography. Then the mechanical pendulum
is etched in a combination of dry and wet etch steps. Alignment of the
etch mask with respect to the metallic leads has to be accurate down to
10~nm in order to provide well defined tunneling contacts. 
In earlier work we
ensured that the processing steps, which are also used in
our present work, provide clean and tuneable tunneling
contacts~\cite{erbe98}. 
An electron micrograph of the measurement setup is shown in
Fig.~\ref{fig1b}.

The first measurements
were performed at room temperature. The sample was mounted in an evacuated
sample holder with a small amount of helium gas added, to ensure
thermal coupling. During evacuation the sample holder was heated. The 300~K trace shows a variety of
resonances, where the source-drain current is increased due to the
motion of the clapper (see Fig.~\ref{fig2b}). 
This behavior is well known from the
measurements performed on the single tunneling barrier~\cite{erbe98}. 
The peaks are
superimposed on a background, which depends linearly on the source-drain
bias. This background is due to the thermal motion of the clapper,
since it disappears at lower temperatures.

The clapper is set into motion by an {\it ac}-voltage applied to the two
driving gates at frequency $f_0 \, $. 
This leads to an alternating 
force acting on the grounded lower part of the pendulum. 
  Additionally, the driving 
voltage also acts as a gate voltage. In order to calculate the
influence of this gate voltage on the number of electrons transferred
we estimate the capacitance between the closest driving electrode and the
island at positions of the island where tunneling occurs. 
This electrode as seen 
by the island has a radius of roughly 215~nm. With a distance of
300~nm 
between driving gate and island, the resulting gate capacitance is
\[
  C \simeq 4 \pi \varepsilon_0 \left(\frac 1 {215 {\rm \, nm}}- \frac 1 {300 {\rm \,nm}}\right)^{-1}
   \approx 84\; \rm aF
\]
which corresponds to gate charges of up to $\pm 527\, e$ if voltages of up to $\pm 1 V$ are applied. 
Because of this large number of
electrons the applied voltages lead to an electrostatic
force on the island. If the frequency $f_0$ of the driving force
coincides with the eigenfrequency of the clapper, resonant motion is
excited. In our calculations this kind of motion shows a large number
of electrons transferred at a high accuracy during each cycle. 
This creates the large peaks seen in the
measurements. If the shuttle moves with frequencies different from 
$f_0$ ({\it e.g.}\/ excited thermally), the resulting current has a
much smaller signal to noise ratio. 
Therefore, the large value of the gate-voltage
explains both the high number of electrons transferred at room
temperature and the background  depending on whether or not the
clapper moves with the same frequency as the applied voltage. 

Transport through the  island on the clapper can be described by a
simple Master equation \cite{weiss99:97} 
\begin{eqnarray} \label{eq:MasterEq}
  \frac{d}{dt} p(m,t) 
  &=& - 
  \left[\Gamma^{(+)}_{\rm L}(m,t)+\Gamma^{(+)}_{\rm R}(m,t)
  \right]p(m,t) \\ 
  &-& \left[ \Gamma^{(-)}_{\rm L}(m,t)+
  \Gamma^{(-)}_{\rm R}(m,t)\right]p(m,t) \nonumber\nopagebreak\\
  &+&\left[\Gamma^{(+)}_{\rm L}(m-1,t)+\Gamma^{(+)}_{\rm R}(m-1,t)\right]p(m-1,t)\nonumber\nopagebreak\\
  &+&\left[\Gamma^{(-)}_{\rm L}(m+1,t)+\Gamma^{(-)}_{\rm
  R}(m+1,t)\right]p(m+1,t)\, \nonumber
\end{eqnarray}
where $\Gamma$ are the transition rates and $p$ the probability to
find $m$ additional electrons on the island at time $t$. 
In a golden rule approach the tunneling rates are of the form~\cite{Lehrbuch}
\begin{equation}
 \Gamma = \frac 1 {e^2R} \, \frac {\Delta E}{1-\exp\left(-\frac{\Delta E}{k_{\rm B}T}\right)}
\, .
\end{equation}
Taking into account that the energy changes $\Delta E$ are time
dependent due to the periodic motion of the shuttle \cite{weiss99:97}
one obtains
\begin{eqnarray}
\Gamma_{{\rm R},m}^{(\mp)}(t) = \hspace{6.3cm}\\ \frac 1{\tau} \; 
  \frac{\pm \left(m +\frac{C_{\rm G}V_{\rm G}}e+ \frac{C_{\rm L}(t)V}{e}\right) -
  \frac 12}{1-\exp\left[-\!\left(\!\pm\!\left(m \!+\!\frac{C_{\rm G}V_{\rm G}}e\!+\! \frac{C_{\rm
  L}(t)V}{e}\right)\!-\! \frac 12\right)\frac{e^2}{C_{\rm \Sigma}(t)k_{\rm
  B}T}\right]} \nonumber \, , 
\end{eqnarray}
where $\tau = R_{\rm R}(t_{\rm max})C_{\rm R}(t_{\rm max})$, 
$t_{\rm max}$ is the time,
where the island is at its closest point to the right electrode, and
$C_{\rm \Sigma}(t)=C_{\rm R}(t)+C_{\rm L}(t) + C_{\rm G}(t)$. 
For the left electrode the indices $R$ and $L$ have to be interchanged, $V$ has to 
be replaced by $-V$ and $x(t)$ by $-x(t)\,$.

Numerical solution of Eq.~(\ref{eq:MasterEq}) results in the
height of the 
peaks being determined by the applied gate--voltage $V_{\rm G}\,$. The number of 
electrons transferred does hardly depend on 
source-drain bias, since the voltages applied on the driving gates
are a factor $10^3$ larger than the source-drain bias which agrees
with the experimental result as shown in the inset of Fig.~\ref{fig2}. 
 The displacement noise of
a cantilever can be calculated in analogy to an electrical
circuit~\cite{yaralioglu99:2379}. 
The mean square displacement is given by the
fluctuation-dissipation-theorem 
$\langle u^2 \rangle = k_{\rm B} T \chi_T$, where $\chi_T = \int {\rm
  d} \omega \, {\rm Im} \chi /\omega$ can be approximated by $\Delta
f \, {\rm Im} \chi / f$ for a peak at frequency $f$ of width $\Delta f$.  
Thus the formula for Johnson noise
derived by Nyquist~\cite{nyquist28:110} can be transformed for the
mechanical case yielding:

\begin{equation} 
        \langle u^2 \rangle = 4 \, k_{\rm B} T \, \Delta f \; {\rm Im} \left\{
         -\frac{1}{\omega} \frac{u_k}{F_i} \right\} ,
\end{equation}     
$u_k$ are the displacements due to applied forces $F_i$. The imaginary part
of $u_i/F_k$ depends on the phase relationship between the actual
motion of the mechanical structure and the force.  
The response of the clapper to an external
force was calculated by a finite element program~\cite{solvia}.
We conclude from this calculation, that contributions to the current
can be expected up to frequencies of some GHz. This explains the
amplitude of the thermal background. 
A contribution to the current by the
substrate can be excluded as well as a displacement current by direct
capacitive coupling of the {\it ac} power by measuring a non-moving
sample~(shown in the inset of Fig.~\ref{fig2}). 

We extracted one peak of the complex spectrum by subtracting the
thermal background and overlapping neighboring peaks.
The
direction of the current in this peak is reversed to the normal
current direction. The peak height does not depend on the {\it
  dc}-source-drain bias as well. Both facts can be explained by the large gate voltage.
In the inset of Fig.~\ref{fig3} the electric field distribution around
the clapper is shown in a top view by assuming a potential $V_{\rm G} =~$1~V at one
of the driving
gates. The calculation was performed by using a finite element program~\cite{mafia}.

The number of transferred electrons in this peak is
approximately 1000 which agrees with our theoretical estimate based on
Eq.~(\ref{eq:MasterEq}). These
results show, that the electron transfer works well at
room-temperature. Since electron tunneling should be very sensitive to
environmental influences this opens an interesting possibility for
sensor applications.

Measurements at lower temperatures show a complete suppression of the
ohmic background and thus indicate its thermal nature. At temperatures
of about 12~K a pronounced peak at 120~MHz is found. The oscillation
amplitude of the motion at the peak position 
is strongly attenuated towards lower temperatures of 4.2~K due to the
increased
stiffness of the clapper. The rest of the complex spectrum is
completely suppressed at 4.2~K also because of the increased
stiffness of the structure.
 The maximum current amplitude of the peak  is (2.3$\pm$0.02)~pA, which
corresponds to
a transfer of 0.11~electrons on average per cycle of motion of the
clapper. Peaks in the classical experiment~\cite{erbe98} show
Lorentzian line shape. 
In order to obtain a formula for the peak fit for the present experiment
we use the simple
expression  
\[
   \left<N\right>
        \propto 
        \frac{t_0} {RC}
\]
for the average number of electrons transferred
derived in \cite{weiss99:97} in the limit of small contact times              
  $t_0 \propto 1/({f\sqrt{x_{\rm max}}})\,$.
The cantilever behaves like a damped harmonic oscillator. 
\begin{equation} \label{eq:harmonisch}
 x''(t) + 2 \pi k x'(t) + 4 \pi^2f_0^2 x(t) =F\sin\left(2 \pi f\,t\right)\,.
\end{equation}
Due to the strong dependence of $R$ on the tunneling distance  
the peak shape can be modelled by following
the approximation:
\begin{equation} 
  N = \frac A {f}\sqrt{\frac{x_{\rm max}(f_{\rm r})}{x_{\rm max}(f)}}\, \exp\left[-B\left\{1\!-\!x_{\rm max}(f)/x_{\rm max}(f_{\rm r})\right\}\right]
\label{eq:peakfit}
\end{equation} 
where
$x_{\rm max}(f) \propto 1/ \sqrt{\left(f^2-f_0^2\right)^2+k^2f^2}$
is the amplitude of the oscillation in resonance and
$f_r = \frac{f_0}{2} \sqrt{4-2k}$
is the shifted frequency of the damped oscillator, respectively. From
the fit parameters we obtain a quality factor \mbox{$Q = {f_0} /k$} of order 10. 
Small $Q$s are essential for operation as a switch where the
oscillating force in Eq.~(\ref{eq:harmonisch}) is replaced by a
step function. For small quality factors the
oscillatory solution of the differential equation vanishes on a short
timescale.

In summary we have shown single electron tunneling 
by using a
combination of nanomechanics and single electron devices. We have
demonstrated a new way to transfer electrons one by one at
radio frequencies. At 4.2~K we measured an average of $0.11\pm 0.001$ electrons which 
shows that the resolution of current transport
through the shuttle should also resolve Coulomb blockade
after minimizing the effects of the driving voltage. 
We estimate the temperature, at which Coulomb blockade should be
observable to be 600~mK. Scaling down the island size will increase
this temperature.  
In future work, forming superconducting and
magnetic islands will be of interest for the understanding of the
tunnel process itself.

We like to thank U.~Sivan for discussions. 
Special thanks to J.~P.~Kotthaus for his support.
This work was financially supported by the Deutsche 
Forschungsgemeinschaft (DFG Bl-487/1-1). CW {\mbox acknowledges} support 
by the SFB~348.


\newpage

\begin{figure}  
\begin{center}
\leavevmode
\caption[fig1b]
{
Electron micrograph of the quantum bell: The pendulum is clamped on
the upper side of the structure. It can be set into motion by an
{\it ac}-power, which is applied to the gates on the left and right hand
side~(G1 and G2) of the clapper~(C).
 Electron transport is then observed from source~(S) to drain~(D) through the
island on top of the clapper. The island is electrically isolated
from the rest of the clapper which is grounded.
}
\label{fig1b}
\end{center}
\end{figure}


\begin{figure} 
\begin{center}
\leavevmode
\caption[fig2b]
{
  Measurement of the tunnel current from source~(S) to drain~(D) at room temperature. 
  The complex spectrum
  is similar to measurements on a single mechanically moving tunnel
  contact \cite{erbe98}. Finite element simulations show a complex
  spectrum of mechanical resonances \cite{solvia}. The inset shows
  a measurement on a non-moving sample. No capacitive crosstalk or
  conduction through the substrate could be observed.  
}
\label{fig2b}
\end{center}
\end{figure}


\begin{figure} 
\begin{center}
\leavevmode
\caption[fig2]
{
Transport through the island is charactreized at a
frequency of 86~MHz. A fit to the peak according to
Eq.~(\ref{eq:peakfit})  
shows differences to the peakshape found in measurements on the single
nanomechanical tunneling contact~\cite{erbe98}. The peak was extracted from the
complete spectrum (shown in Fig.~\ref{fig2b}) by subtracting the thermal
background and overlapping neighboring peaks. The inset shows
the peak at source-drain voltages from 0~mV (lowest curve) to 10~mV
(top curve). 
}
\label{fig2}
\end{center}
\end{figure}

\begin{figure} 
\begin{center}
\leavevmode
\caption[fig3]
{
At low temperatures only one peak remains, which can be
fitted by the model given in Eq.~(\ref{eq:peakfit}). Inset:
Capacitive cross-coupling between the driving gates and the island
calculated by a finite element program~\cite{mafia}. 
}
\label{fig3}
\end{center}
\end{figure}


\end{document}